\documentclass[prb,twocolumn,showpacs,amsmath,amssymb,superscriptaddress]{revtex4}
\usepackage{graphicx}
\usepackage{dcolumn}
\usepackage{bm}
\usepackage{epsfig}

\begin{document}
\title{Transverse field muon-spin rotation signature of the skyrmion
  lattice phase in Cu$_{2}$OSeO$_{3}$}
\author{T. Lancaster}
\email{tom.lancaster@durham.ac.uk}
\affiliation{Centre for Materials Physics, Durham University, Durham,
  DH1 3LE, United Kingdom}
\author{R.C. Williams}
\affiliation{Centre for Materials Physics, Durham University, Durham,
  DH1 3LE, United Kingdom}
\author{I.O. Thomas}
\affiliation{Centre for Materials Physics, Durham University, Durham,
  DH1 3LE, United Kingdom}
\author{F. Xiao}
\affiliation{Centre for Materials Physics, Durham University, Durham,
  DH1 3LE, United Kingdom}
\author{F.L. Pratt}
\affiliation{ISIS Facility, STFC Rutherford Appleton Laboratory,
  Chilton, Didcot, Oxfordshire, OX11 0QX, United Kingdom}
\author{S.J. Blundell}
\affiliation{Oxford University Department of Physics, Clarendon Laboratory, Parks
Road, Oxford, OX1 3PU, United Kingdom}
\author{J.C. Loudon}
\affiliation{Department of Materials Science and Metallurgy, University of Cambridge, 27 Charles Babbage Road,
Cambridge CB3 0FS, United Kingdom}
\author{T. Hesjedal}
\affiliation{Oxford University Department of Physics, Clarendon Laboratory, Parks
Road, Oxford, OX1 3PU, United Kingdom}
\author{S.J. Clark}
\affiliation{Centre for Materials Physics, Durham University, Durham,
  DH1 3LE, United Kingdom}
\author{P.D. Hatton}
\affiliation{Centre for Materials Physics, Durham University, Durham,
  DH1 3LE, United Kingdom}
\author{M. Ciomaga Hatnean}
\author{D.S. Keeble}
\author{G. Balakrishnan}
\affiliation{University of Warwick, Department of Physics, Coventry,
  CV4 7AL, United Kingdom}

\begin{abstract}
We present the results of transverse field (TF) muon-spin rotation ($\mu^{+}$SR)
measurements on  Cu$_{2}$OSeO$_{3}$, which has a skyrmion lattice (SL)
phase.
We measure the response of the TF $\mu^{+}$SR signal in that phase
along with 
the surrounding ones, and suggest how the phases might be
distinguished using the results of these measurements. 
Dipole field simulations support the conclusion that the muon
is sensitive to the SL via the TF lineshape and, 
based on this interpretation, our measurements suggest that the SL is
 quasistatic on a timescale $\tau > 100$~ns. 
\end{abstract}
\pacs{12.39.Dc, 76.75.+i, 74.25.Uv}
\maketitle

\section{Introduction}

The understanding of matter and its excitations in terms of topology has a long history which is now allowing
us an insight into the properties of quantum materials and the opportunity for their manipulation.
Topological physics includes the study of kinks,
vortices and monopoles in quantum field theory, which forms the basis of a successful branch of
condensed matter physics \cite{chaikin,nagaosa}. 
A skyrmion \cite{skyrm} is a topological object which has been shown to exist in a range of magnetic materials at particularly well-defined conditions of temperature and magnetic field.
The simplest example of a skyrmion  may be derived from a sphere
studded with arrows pointing radially. A skyrmion is formed if
we stereographically project the arrows onto a plane while keeping their orientations
fixed. This produces a twisting pattern which turns out to be
a stable knot; it cannot be untied as long as the fields remain smooth
and finite. While skyrmions exist in different forms in a variety
of physical systems, the clearest evidence for the existence of the skyrmion is in the spin texture of magnetic systems and in
recent years a number of spectacular advances have demonstrated  not only the existence of magnetic
skyrmions, but also their ordering into a skyrmion lattice (SL)
\cite{Muhlbauer-2009,Munzer-2010,Yu-2010,Yi-2010a,Seki-2012a,Seki-2012b,Adams-2012,Seki-2012c,Langner-2014,omrani}.
There is considerable similarity between the SL, which has recently
been shown to be present in a range of
non-centrosymmetric, helimagnetic systems, and another topological phase: the vortex lattice
(VL) found in type II superconductors in applied magnetic field. Like the SL, the superconducting VL leads to
a textured, periodic distribution of magnetic field within the body of the material, which is large on the scale
of the crystallographic unit cell. In practice, the same experimental techniques that have been successfully
applied to probing VL physics, such as small angle neutron scattering \cite{forgan}
and, in particular, muon-spin rotation
($\mu^{+}$SR) \cite{steve,sonier}, could potentially be
applied to probe the physics of the SL and this hypothesis forms the basis of the
work reported here.
 In this
paper
we investigate the extent to which  transverse field (TF)
$\mu^{+}$SR measurements are sensitive to the skyrmion phase in the multiferroic skyrmion material
Cu$_{2}$OSeO$_{3}$. 

\begin{figure}
\begin{center}
\epsfig{file=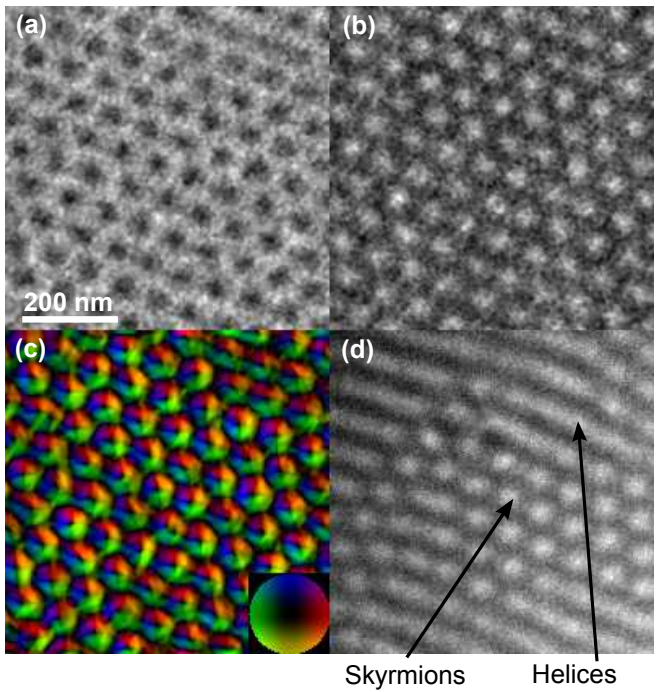,width=\columnwidth}
\caption{
(a),(b) Skyrmions in Cu$_{2}$OSeO$_{3}$ imaged with Lorentz TEM at 10~K in a
field of 30~mT with
defoci $\pm0.7$~mm. 
(c) Magnetic induction map constructed from (a) and
(b) using the transport of intensity equation. The direction of the
local induction is indicated by the colour wheel (inset). 
(d)
Skyrmions coexisting with the helical phase at 30~K in a field of 40~mT.
Similar observations were made by Seki et al.\cite{Seki-2012a} 
\label{fig:LTEM}}
\end{center}
\end{figure}

Our use of TF $\mu^{+}$SR to probe the SL in Cu$_{2}$OSeO$_{3}$ is analogous to its use in probing the VL in
a type II superconductor, where the technique provides a powerful
means of measuring
the internal magnetic
field distribution caused by the presence of the magnetic field
texture \cite{steve,sonier}.
In a TF $\mu^{+}$SR experiment, spin polarized muons are implanted
in the bulk of a material, in the presence of a
magnetic field $B_{\mathrm{a}}$, which is applied perpendicular
to the initial muon spin direction. Muons stop at
random positions on the length scale of the field texture where they
precess about the total local magnetic field $B$ at the muon
site (mainly due to the field texture), with frequency 
$\omega = \gamma_{\mu}B$, where $\gamma_{\mu} = 2 \pi \times
135.5$~MHz T$^{-1}$.
The observed property
of the experiment is the time evolution of the muon
spin polarization $P_{x}(t)$, which allows the determination of
the distribution $p(B)$
 of local magnetic fields across the
sample volume via $P_{x}(t) = \int_{0}^{\infty}\mathrm{d}B\, p(B) \cos
(\gamma_{\mu}Bt + \phi)$
where the phase $\phi$ results from the detector geometry. 

For our measurements powder and single crystal samples of  Cu$_{2}$OSeO$_{3}$ were synthesized as described
previously \cite{Bos-2008}, with the same batch of polycrystalline powder also used for the
growth of single crystals using the chemical vapour transport technique.
The samples were characterised with powder and single crystal x-ray diffraction which confirmed that the samples were of the requisite B20 phase. 
The magnetic response was checked with magnetic susceptibility
measurements which confirmed the magnetic transition observed at $T=57$~K.

To establish the existence of skyrmions in our Cu$_{2}$OSeO$_{3}$ samples,
a sample was prepared for electron microscopy. Single crystal
samples showed clear facets and the sample was mechanically polished
on a (110) face until it was 20~$\mu$m thick. It was then further thinned
by argon ion beam irradiation using a Gatan Precision Ion Polishing
System (PIPS) initially operated at 4~kV with the ion guns set at $7^{\circ}$ to the sample plane. Once the sample thickness approached the
wavelength of light, rainbow colours could be observed using an
optical microscope and, at this point, the voltage was reduced to 2~kV
and the gun angle to $5^{\circ}$ and thinning continued until a hole of
size 15~$\mu$m appeared. Images were taken from an area of sample
approximately 50~nm thick surrounding this hole with an FEI Tecnai F20
transmission electron microscope (TEM) equipped with a field-emission gun
using an acceleration voltage of 200~kV. In normal operation, the
objective lens of the microscope applies a 2~T field to the specimen
which would force it into its ferrimagnetic state so images were
acquired in low-magnification mode in which the image is formed using
the diffraction lens and the objective lens was weakly excited to
apply a small magnetic field normal to the plane of the specimen. The
sample was cooled using a Gatan liquid-helium cooled
`IKHCHDT3010-special' tilt-rotate holder which has a base temperature
of 10~K. The images were recorded on a CCD using a Gatan Imaging
Filter and they were energy-filtered so that only electrons which had
lost between 0 and 1~ eV on passing through the specimen contributed
to the image. An aperture was also used to ensure that only the
000-beam and the low-angle scattering from the skyrmions contributed
to the image. The defocus and magnification were calibrated by
acquiring images with the same lens settings from Agar Scientific’s
`S106' calibration specimen, which consists of lines spaced by 463~nm
ruled on an amorphous film. The defocus was found by taking digital
Fourier transforms of the images acquired from the calibration
specimen and measuring the radii of the dark rings that result from
the contrast transfer function as previously described\cite{williams}. 
These TEM measurements on the thinned crystallites confirm the presence of
skyrmion spin textures, as shown in Fig.~\ref{fig:LTEM}. These occur
over an extended region of the phase diagram, compared to results on
 bulk materials, as observed previously \cite{Seki-2012a}. We also
 note the coexistence of skyrmions and helices, as seen in the
 previous microscopy study\cite{Seki-2012a}.

TF $\mu^{+}$SR measurements
were made 
using the MuSR spectrometer at the ISIS facility and the GPS
spectrometer at the Swiss Muon Source (S$\mu$S). For measurements at
ISIS the sample was mounted on a hematite backing plate in order that
muons that are implanted in the hematite would be rapidly depolarized
and therefore removed from the spectrum. For measurements at GPS the
so-called flypast geometry was employed, where the powder sample was suspended
on a fork, which prevents those muons that do not implant in the
sample from contributing to the signal. For all measurements
presented, the
sample was cooled from $T>60~K$ in the applied field.
All data analysis was carried out using the WiMDA program
\cite{WiMDA}.

\section{Results of TF $\mu^{+}$SR measurements}

\begin{figure*}
\begin{center}
\epsfig{file=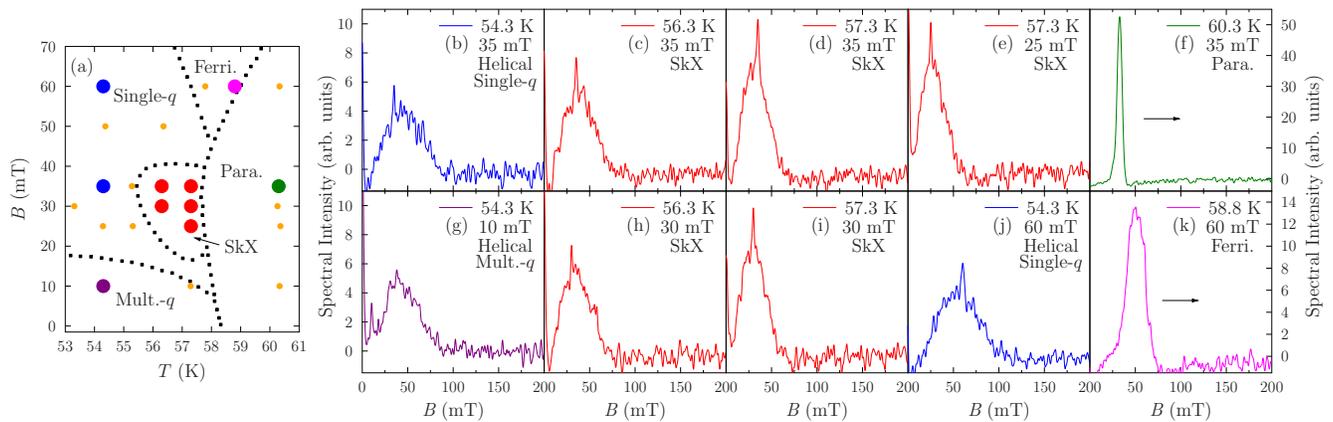,width=17.5cm}
\caption{
(a) The phase diagram of Cu$_{2}$OSeO$_{3}$, indicating the locations of the
  paramagnetic, ferrimagnetic, single-$q$ and
  multiple-$q$ helical phases.
 Small circles show where measurements were taken, 
and the bold circles correspond to the frequency spectra
displayed in (b-k). 
These phase boundaries are taken from 
Ref~\onlinecite{Seki-2012a}.
(b-k) Evolution of the magnetic field distribution measured at
S$\mu$S, with
  temperature and applied field. (b,j) Single-$q$ helical
  phase; (c-e, h,i) the SkX phase containing the SL; (f)
  paramagnetic phase; (g) the multi-$q$ helical phase; and (k) the
  Ferrimagnetic phase. 
All data are plotted to the same spectral intensity 
scale, except (f) and (k).
 \label{fig:freq-domain}}
\end{center}
\end{figure*}

The region of the  Cu$_{2}$OSeO$_{3}$ phase diagram of interest in
this study is shown in Fig.~\ref{fig:freq-domain}(a) 
We note that this is the phase diagram from Ref~\onlinecite{Seki-2012a},
which differs slightly from that derived from other studies
(e.g. Refs~\onlinecite{omrani, Adams-2012}). 
In this material, spins are located on the Cu$^{2+}$ ions, which are arranged
in four tetrahedra within the unit cell. Ferrimagnetic (FiM) ordering  of
these spins is favoured, with one Cu ion (Cu1) oriented in the opposite direction to that of
the others  (Cu2) \cite{Bos-2008,Maisuradze-2011}.  This ordered configuration
persists when  we build  more complicated spin  configurations
described below
\cite{Janson-2014}.
In systems such as Cu$_{2}$OSeO$_{3}$ which  lack inversion symmetry,
chiral interactions are capable
of stabilising  helically ordered configurations of  spins
\cite{Rossler-2006}.   
An applied external magnetic field  can further
stabilise a superposition of three helical configurations in the plane
perpendicular  to  the  field (the  configuration  is  translationally
invariant  in the  remaining  direction).
The skyrmion  spin textures can be formed from a triangular arrangement of
the  wave vectors of  these helices \cite{Petrova-2010}. 
From the Cu$_{2}$OSeO$_{3}$  phase diagram, we see that below a lower critical field a phase described as
the multi-$q$ helimagnetic (or simply the helimagnetic phase)
exists.   In bulk  single-crystal samples,  this consists  of multiple
domains of  helimagnetic stripes whose  $q$ vectors are  aligned along
the       three      $\langle       001      \rangle$       directions
\cite{Seki-2012b,Adams-2012}.
Above this lower critical field  but below a second, larger critical
field  (above which  there is a transition to  a ferrimagnetic phase),
the $q$  vector is  aligned parallel  to the  direction of  the field,
forming a  single-$q$ helimagnetic  phase (also called the conical phase)
where        there        are        no        multiple        domains
\cite{Seki-2012b,Adams-2012,Seki-2012a}.  
In  Cu$_{2}$OSeO$_{3}$  the A-phase or
skyrmion lattice phase (often denoted SkX) is centered around $B_{\mathrm{a}}=30$~mT at $T=57$~K,
close to the critical temperature $T_{\mathrm{c}}$  where the system undergoes a  transition to  a paramagnetic
state \cite{Seki-2012b,Adams-2012,Seki-2012a}.
The SkX phase may be described by a phase-locked
superposition  of   three  helimagnetic  textures  with   $q$  vectors
perpendicular  to the  direction  of  the field,  such  that a  regular
lattice of skyrmions  is formed.  It has further been  shown in a bulk
single crystal system  that this phase may be  further subdivided into
two subphases,  with the lattice in  the second being rotated  by thirty
degrees with respect to the lattice in the first \cite{Seki-2012b}.
The range of applied magnetic fields over which the SkX phase occurs
in Cu$_{2}$OSeO$_{3}$  makes it
 well matched to the ISIS and the S$\mu$S muon source time windows.

\begin{figure}
\begin{center}
\epsfig{file=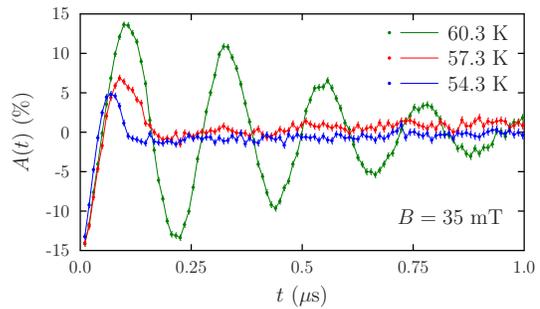,width=7cm}
\caption{(a) Time domain spectra measured in an applied transverse
  magnetic field of 35~mT in the paramagnetic phase at
  60.3~K, the skyrmion phase at 57.3~K and the single-$q$ phase at 54.3~K. \label{fig:timedomain}}
\end{center}
\end{figure}

\begin{figure}
\begin{center}
\epsfig{file=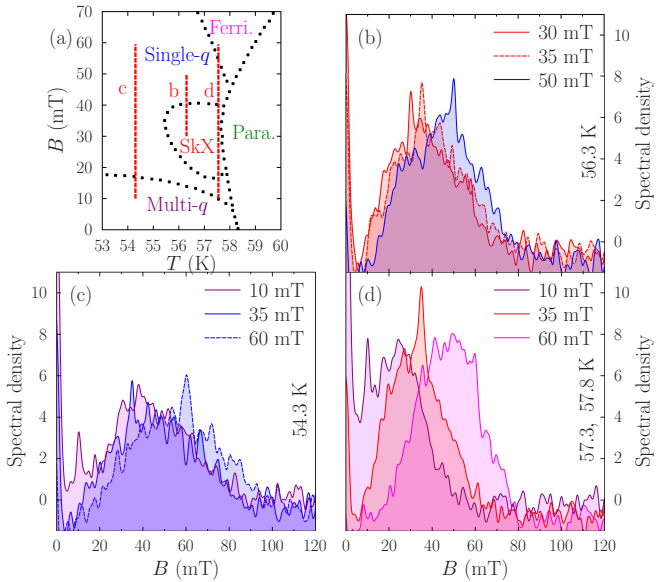,width=\columnwidth}
\caption{(a) The cuts through the phase diagram at constant temperature displayed in (b-d). 
(b)  Applied magnetic field-dependence of the TF lineshape 
  measured at $T=56.3$~K. (c) Field sweep within the multi- and single-$q$ phases
  at 54.3~K. 
(d)  Field sweep through multi-$q$, SkX and ferrimagnetic phases. 
\label{fig:comparison}}
\end{center}
\end{figure}

Example TF $\mu^{+}$SR time-domain spectra  measured at S$\mu$S are shown in
Fig.~\ref{fig:timedomain}, where the increase in damping upon entering
a magnetically ordered phases (in this case the skyrmion and single $q$-phases) is
evident. For our purposes, it is more instructive to consider the
frequency domain spectra, as shown in
Figs.~\ref{fig:freq-domain} and \ref{fig:comparison}.
These were obtained via a fast Fourier transform of the time-domain
spectra, using a Lorentzian filter with time constant 2.2~$\mu$s.  
[In these spectra the spectral density is proportional to $p(B)$.]

We see that spectra in the skyrmion phase are characterized via a
relatively narrow, asymmetric spectral density distribution, which
rises steeply on the low field side leading to a sharp peak, when
compared to spectra within the helical phases. This peak then decays,
first very rapidly and then more gradually with increasing field. In
contrast, the single- and multi-$q$ helical phases give rise to less
asymmetric distributions with broader maxima and spectral weight
distributed over a larger range of fields.
In further contrast, the ferrimagnetic phase shows a narrowed, fairly
symmetric line; while  the paramagnetic regime is
characterized by a very narrow linewidth (note the change of scale in
Fig.~\ref{fig:freq-domain} for spectra from these latter two phases).

Spectra in the single-$q$ phase do not show a significant change
with temperature  below 55~K. 
However,  as we approach temperatures close to
the transition to the paramagnetic region
[Fig.~\ref{fig:comparison}], 
we see a temperature-driven effect  in the single-$q$, multi-$q$ and
skymion regions, which
causes  the spectra 
to show more spectral weight at the applied
field, taking on a more heavily peaked appearance. 
This effect is evident in several spectra
[see e.g.\ Fig.~\ref{fig:comparison}(d)], 
but can be most clearly seen in the single-$q$ phase data measured at
50~mT and 56.3~K[Fig.~\ref{fig:comparison}(b)] when compared to
spectra measured at lower temperature in that phase
[Fig.~\ref{fig:comparison}(c)], which are quite unresponsive to
changes in applied field.
 However, comparing this latter single-$q$ phase  spectrum  with those measured at
 the same temperature, but in fields thought to promote the SkX phase [Fig.~\ref{fig:comparison}(b)], we
 see that the spectra may still be distinguished: the SkX-phase spectra
are more asymmetric, with  spectral weight persisting on the high
field side of the peak. 
We note that in the Lorentz TEM measurements there is evidence for the
coexistence of skyrmions and helical spin textures
[Fig.~\ref{fig:LTEM}(d)]. 
It could be, therefore, the case that the
data point in question ($B=50$~mT and $T=$56.3~K), measured on a polycrystalline sample,
features a contribution from both single-$q$ and SkX phases. However,
it is also worth recalling that the phase diagram of the thinned samples
supports the existence of skyrmions over a wider range of the $B$-$T$
phase diagram than for the bulk, so the cases are not
straightforwardly comparable. 

\begin{figure}
\begin{center}
\epsfig{file=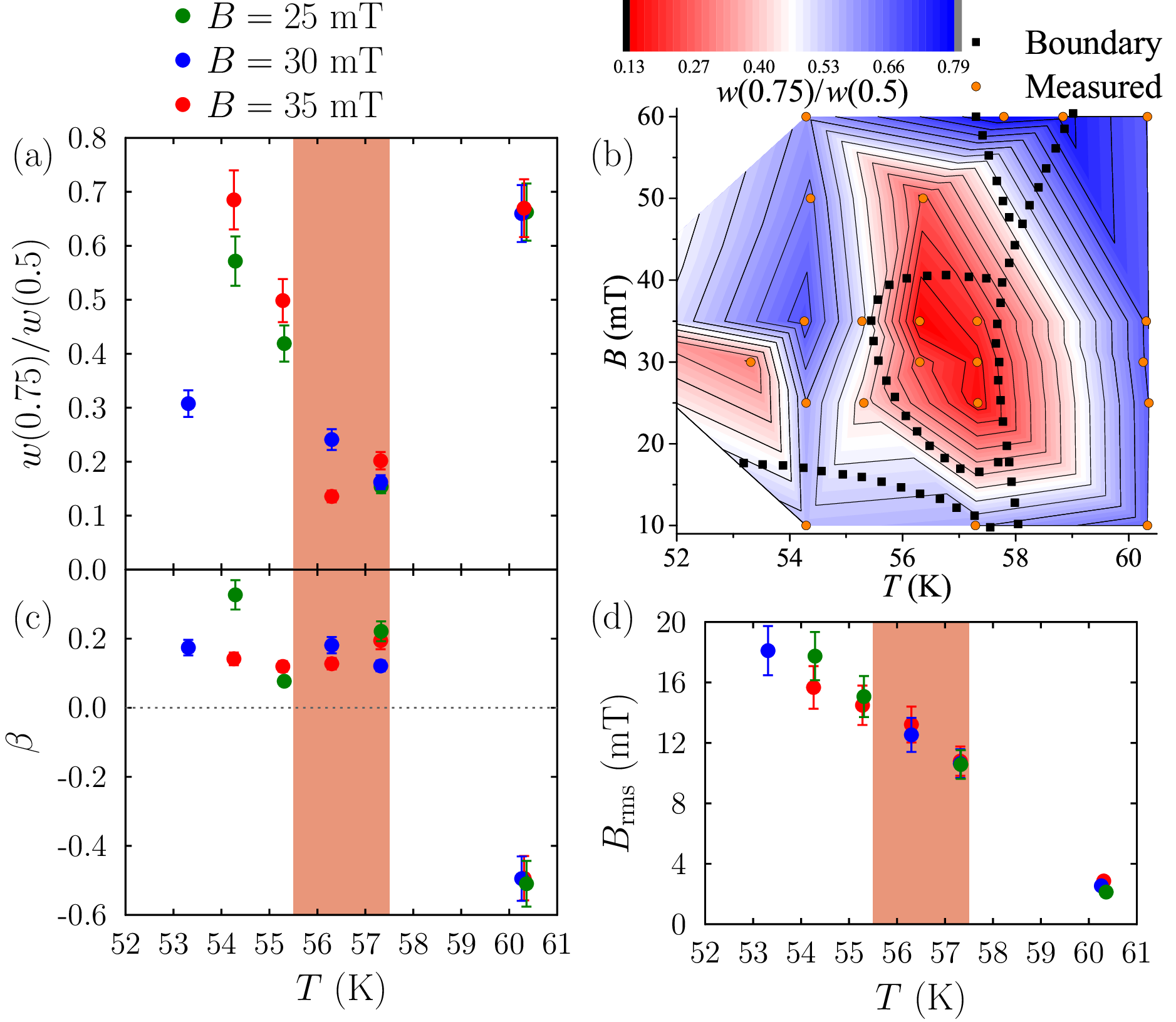,width=\columnwidth}
\caption{(a) The ratio of peak widths $w(0.75)/w(0.5)$ discussed in the
  text. (b) Contour plot of $w(0.75)/w(0.5)$ from which the SkX phase
  may be identified. 
(c) The skewness  $\beta$ and (d) standard deviation $B_{\mathrm{rms}}$ as
a function of temperature. 
 \label{contour}}
\end{center}
\end{figure}

One crude way to identify the SkX region of the phase diagram from
these data is via the relatively
tapered lineshape observed in that phase. 
To this end, Fig.~\ref{contour}(a) shows 
one possible parametrization of our data that approximately captures the
characteristic signal in the SkX region. 
In Fig.~\ref{contour}(a) we show the
ratio of $w(0.75)$ (the full peak width at 0.75 of the maximum spectral
weight) to $w(0.5)$ (the full width at half maximum). The contour plot
in Fig.~\ref{contour}(b) shows that this quantity picks out the SkX phase as an island centered around 57~K
and 30~mT where $w(0.75)/w(0.5)$ is suppressed. 
Moreover, despite the resolution of our contour plot
being limited by the number of data points, we note the resemblance to
the Skyrmion phase boundaries reported elsewhere. 
Most of the spectra obtained outside of the paramagnetic region
display a small additional peak at the value of the applied field, which 
we ascribe to the diamagnetic response from a small portion of the muon 
ensemble. This might cause a further suppression of $w(0.75)/w(0.5)$
within the SkX phase, whose suppression is otherwise caused by 
the tapered nature of the lineshape, as the
diamagnetic contribution coincides with the peak in spectral weight 
 in the muon response caused by the internal field distribution. 
The limitation of this
method is apparent, however, in that the data point  at 50~mT and
56.3~K 
discussed above gives a rather similar
ratio to the points in the SkX region, despite its lineshape being
distinguishable via its asymmetry. 
As described above, one interpretation of this spectrum
 involves  some portion of the signal arising
from skyrmions, but it remains distinct from the other spectra from
the SkX phase, suggesting that there is a significant single-$q$ phase
contribution.

The normalized skewness of the distribution $\beta$, defined via 
$\beta = 	\int {\rm d}B ~ p(B) \left[ \frac{ (B- \langle B \rangle)}{B_{\rm rms}}
 \right]^3$ is shown in Fig.~\ref{contour}(c).
We find that $\beta$ is slightly enhanced in the SkX phase compared to the single-$q$ phase,
before dropping to a negative value in the paramagnetic (PM) phase. 
Finally, the standard deviation Fig.~\ref{contour}(d) of the distribution $B_{\mathrm{rms}} =
\sqrt {\langle (B - \langle B \rangle )^{2} \rangle}$ 
increases fairly smoothly with decreasing temperature, but is
 less sensitive to the SkX phase boundaries. 

Our results suggest that TF $\mu^{+}$SR is sensitive to the
SkX phase through the muon line shape. 
Assuming that this is the case we may consider the mechanisms for muon relaxation and the timescales
involved. 
We note that, in contrast to muon spectroscopy, neutron measurements
of the SL are insensitive to 
fluctuations on timescales much slower than $\hbar/\Delta E \approx10^{-11}$s (where 
$\Delta E \approx 1$~meV is the energy scale of the resolution of the
measurement)
and so fluctuations on timescales longer than this appear static.
Neutrons therefore take a ``snap-shot'' of the behaviour
compared to $\mu^{+}$SR measurements whose 
characteristic time scale is set by the gyromagnetic ratio
of the muon ($\gamma_{\mu} = 2 \pi\times 135.5$~MHz
T$^{-1}$). 
The sensitivity of the
muon to the SL would imply that the signal from the 
long-range spin texture
 is not  significantly motionally narrowed. 
This would imply that the SL itself is
likely to be quasistatic on timescales $\tau \gg (1/\gamma_{\mu}
B_{\mathrm{rms}}) > 100$~ns. This is consistent with the
long lengthscale dynamics observed via Lorentz microscopy \cite{mochizuki},
which suggest rotations of the skyrmion texture take place on  a
0.1~s timescale. 
Consistent with the observation that the muon signal below
$T_{\mathrm{c}}$ is dominated by the the distribution of static internal
magnetic fields are the results of
the previous zero-field $\mu^{+}$SR study of this
material\cite{Maisuradze-2011}. 
Those measurements showed sizeable
transverse relaxation of the muon precession signal, which decreased
as the transition was approached from below, consistent with the
  decrease of the internal field with increasing temperature. This was
  accompanied by an increase in a small longitudinal
  relaxation rate, suggesting a slowing
of residual dynamic fluctuations above 50~K (i.e.\ in the region of interest
for our study). 
It might be expected that the nature of the dynamics changes in each
of the magnetic phases. However, the contribution to the muon
linewidth from the dynamic effects would seem to make only a small
contribution to the overall shape.

\begin{figure}
\begin{center}
\epsfig{file=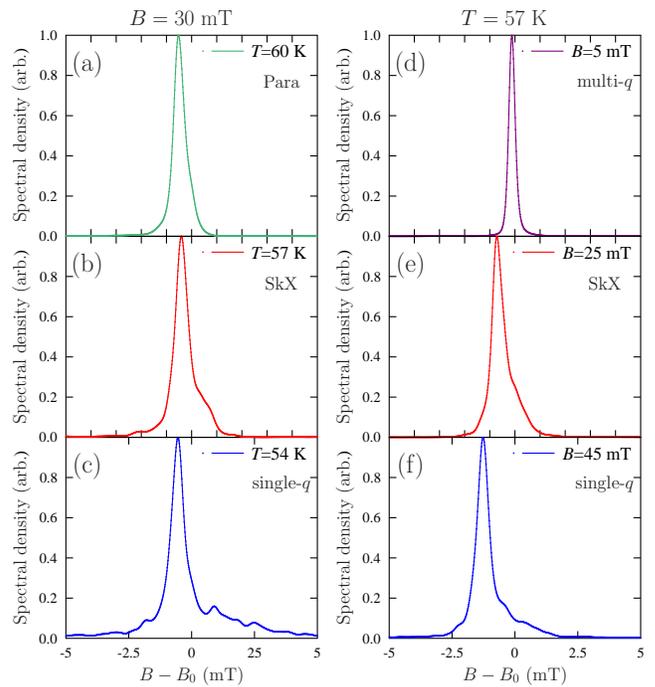,width=\columnwidth}
\caption{Evolution of the magnetic field distribution (measured at
  ISIS) for measurements made on a powder sample in a field of 30~mT (a-c) and  at a
  temperature of 57~K (d-f).
 (a) Paramagnetic phase; (b,e) The SkX phase containing
the SL; (c,f) single-$q$ helical phase; and (d) multi-$q$ phase.\label{fig:isis}}
\end{center}
\end{figure}

Further TF measurements were also made on a powder sample
at ISIS using the MuSR instrument. In this case, we obtain
 similar responses as those reported above, although the analysis of
these are complicated by a sizeable background
contribution from muons stopping in the sample holder or cryostat
tails. Example frequency spectra, obtained via maximum entropy analysis of the
measured time-domain spectra, are shown in Fig.~\ref{fig:isis} for
cuts through the SkX phase in field and temperature. 
 In
these we see that a knee in the spectrum develops in the SkX phase
Fig.~\ref{fig:isis}(b,e) on the high
field side of the peak region [Fig.~\ref{fig:isis}(a)]. This
corresponds to the slow decay of the lineshape in high field seen in
the S$\mu$S data in Fig.~\ref{fig:freq-domain}. This feature is lost
when we enter the single-$q$ phase Fig.~\ref{fig:isis}(c,f),
where the spectral weight in the knee is seen to decay into a broad
distribution. In the multi-$q$ phase, the line is seen to be very
narrow, suggesting that the background contribution is dominant in
this case. 

Measurements were repeated on
 other batches of powder
sample, which showed some slight deviations in the position of the
phase boundary to the paramagnetic phase,
which we attribute to sample variability. We also made 
measurements at ISIS on mosaics of unoriented single crystal samples, which showed
consistent results, although the lack of angular averaging and some
sample variability makes a detailed comparison difficult.

\section{Discussion}

In order to understand these results we have simulated the expected 
distribution $p(B)$ for the spin configurations of this material. 
Owing to the complexity of the problem in a system with multiple muon
sites and a complicated spin configuration,  we have simulated the
field distributions expected for a single crystal orientation, with
the applied magnetic field along $[001]$. For simplicity we generate
our skyrmion spin textures here using the three-$q$ approach outlined
below. We note that more sophisticated treatments of the SL are possible
(e.g.~Ref.~\onlinecite{bogdanov}),  whose predictions are generally in better
agreement with observations. 
For our simulation, skyrmion-like  spin  configurations  \cite{Muhlbauer-2009,Seki-2012c}  were  generated  using
$\boldsymbol{m}(\boldsymbol{r}) = \boldsymbol{m}_{\mathrm{sk}}(\boldsymbol{r})/|\boldsymbol{m}_{\mathrm{sk}}(\boldsymbol{r})|$ 
where
\begin{equation}
  \boldsymbol{m}_{\mathrm{sk}}(\boldsymbol{r})=\sum_{i=1}^{3}
\left[\hat{\boldsymbol{e}}_{z}
    \cos(\boldsymbol{q}_{i}\cdot \boldsymbol{r}+\pi)+
    \hat{\boldsymbol{e}}_{i}
\sin(\boldsymbol{q}_{i}\cdot \boldsymbol{r}+\pi)\right],
\end{equation}
where  $\boldsymbol{q}_{i}$  are the  skyrmion  lattice  modulation
vectors, taken to be perpendicular to the applied field
$B_{\mathrm{a}}$ ($\parallel \hat{\boldsymbol{e}}_{z}$), and $\hat{\boldsymbol{e}}_i$ are the
unit vectors of the skyrmion lattice (also perpendicular to $B_{\mathrm{a}}$).
We generate a 
skyrmion lattice magnetization texture in the [110] plane by taking
$\boldsymbol{q}_{1}=F(-1,\,0,\,0)$,                   
$\boldsymbol{q}_{2}=F(\frac{1}{2},\,-\frac{\sqrt{3}}{2},\,0)$,             
$\boldsymbol{q}_{3}=F(\frac{1}{2},\,\frac{\sqrt{3}}{2},\,0)$,     
$\hat{\boldsymbol{e}}_{1}=(0,\,1,\,0)$,     
$\hat{\boldsymbol{e}}_{2}=(-\frac{\sqrt{3}}{2},\,-\frac{1}{2},\,0)$
and             
$\hat{\boldsymbol{e}}_{3}=(\frac{\sqrt{3}}{2},\,-\frac{1}{2},\,0)$,
where $F=2\pi/L_{\mathrm{sk}}$  and $L_{\mathrm{sk}}$  is the
skyrmion wavelength.
For our simulations we used a value of $L_{\mathrm{sk}}=70$ unit
cells, similar to that suggested in Ref.~\cite{Seki-2012b}.
This method results in a ferromagnetic  (FM)
ordering   of   the    spins within a crystalline unit cell.   
By reversing the  spin  on  the  Cu1 sites, we generate
ferrimagnetic  (FiM)  configurations,   which  results in textures
that match  the
observed magnetic structures of the system \cite{Bos-2008}.

In addition to the SL, we have generated spin textures for the helical
phases. 
Single-$q$ helical configurations were calculated for $B_{\mathrm{a}}$ 
parallel
to  the [001] direction 
using  \cite{Seki-2012c}:
\begin{equation}
\boldsymbol{m}(\boldsymbol{r})=-
\frac{1}{2}\left[\hat{\boldsymbol{e}}_{x}
\cos(q_z z)    +
\hat{\boldsymbol{e}}_{y}\sin(q_z z)\right],
\end{equation}
where $q_z=2\pi/L_{sk}$. Reversing the Cu1 spin direction again
generates the FiM
configurations.  
The exact spin configuration of the more complicated multi-$q$ phase 
is not confirmed and so we have investigated a trial structure.  
We chose to  model this phase by calculating the field
distributions  for FiM  configurations  reflecting six possible  domains,
each with equal population.  Each domain  has $\boldsymbol{q}$
either parallel or  
antiparallel to
the $[100]$, $[010]$  or $[001]$ axes. Its texture is
given by
$\boldsymbol{m}(\boldsymbol{r})=
\frac{1}{2}\left[ \hat{\boldsymbol{e}}_i
\cos(\boldsymbol{q}\cdot\boldsymbol{r})    +
\hat{\boldsymbol{e}}_j\sin(\boldsymbol{q}\cdot\boldsymbol{r})\right] 
$
where $\hat{\boldsymbol{e}}_i$ and $\hat{\boldsymbol{e}}_j$  
are unit vectors in  the plane
perpendicular  to   $\boldsymbol{q}$  and
$\boldsymbol{q}=\pm \hat{\boldsymbol{e}}_{q}|\boldsymbol{q}|$ 
for the parallel and anti-parallel cases respectively. Here
$\hat{\boldsymbol{e}}_{q}$ is the unit vector in the direction of $\boldsymbol{q}$
and 
$|\boldsymbol{q}|=2\pi/L_{\mathrm{sk}}$.
 We   average   over   the
distributions for  the six resulting  textures in order to  obtain our
overall distribution.

We assume muons couple to the dipolar magnetic fields of the Cu$^{2+}$ spins in
the material. 
The dipole magnetic  field  component $B^{i}$  at a position $\boldsymbol{r}$ are    given   by
\cite{steve,Maisuradze-2011}:
\begin{equation}
B^{i}(\boldsymbol{r})=
\frac{C\mu_0}{4\pi}
\sum_{n,j}\frac{m^j(\boldsymbol{r}_{n})}{R_n^{3}}
\left(\frac{3 R_{n}^{i}R_{n}^j}{R_{n}^{2}}-\delta^{ij}\right) 
+ B_{\mathrm{a}},
\end{equation}
 where $i$ and $j$ run  over all three cartesian
directions, $n$ labels
a Cu ion at position $\boldsymbol{r}_{n}$, $C=0.25/\sqrt{3}$ scales  the magnetic
moment to match the observed Cu ions in  these systems (see below)
and $\boldsymbol{R}_{n}=\boldsymbol{r}-\boldsymbol{r}_{n}$.
A previous zero field $\mu^{+}$SR study of this material \cite{Maisuradze-2011}  found five muon  stopping sites
in each crystalline unit cell at positions $A = (0.215, 0.700,  0.970), B = (0.035,
0.720, 0.805),  C = (0.195, 0.555,  0.795), D = (0.275,  0.295, 0.460)$
and $E  = (0.635, 0.550,  0.525)$.  
We have simulated the  field distribution $p(B)$ expected in each
of the phases at a variety of applied magnetic fields, assuming
these previously proposed muon sites. For these simulations we take the
magnetic moment to be $0.25 \mu_{\mathrm{B}}$, consistent with that
found in a previous study \cite{Bos-2008}. For comparison with the
measured spectra, the distributions were convolved with a simulated
instrument function, generated with the WiMDA program.

\begin{figure*}
\begin{center}
\epsfig{file=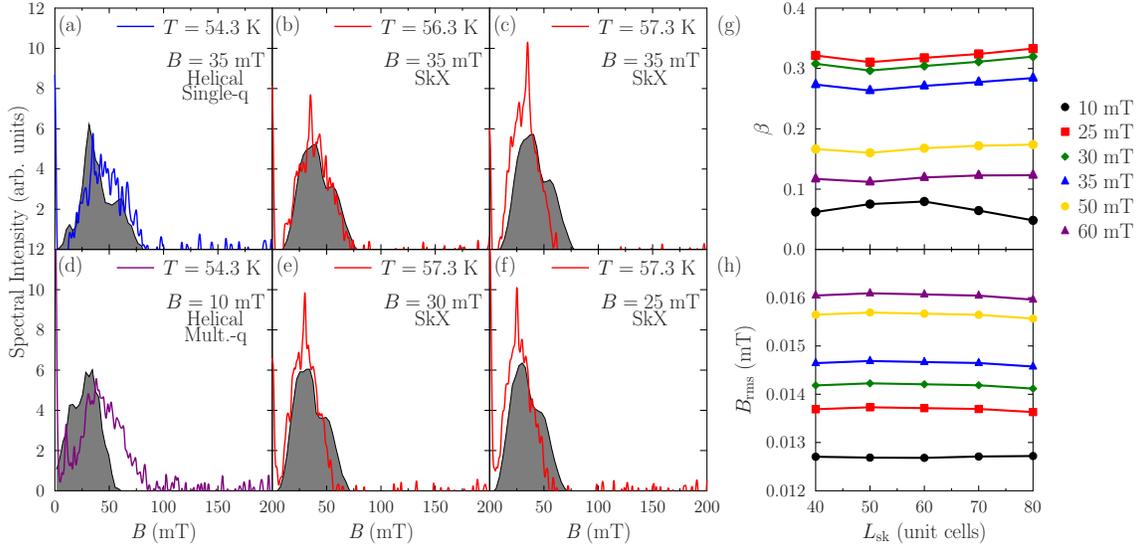,width=15cm}
\caption{(a-f) Results of dipole simulations of $p(B)$ in each phase
  (shown shaded)
  compared with the spectra measured at S$\mu$S. In order to compare the
  simulation and experiment, the area
  under the simulated distribution is scaled to that under the
  experimental distribution, assuming a constant background contribution. 
Scaling of the (g) skewness and (h) $B_{\mathrm{rms}}$ of the
simulated distribution
with the skyrmion lattice size $L_{\mathrm{sk}}. $
\label{fig:simulation}}
\end{center}
\end{figure*}

The results of these simulations are compared with the observed spectra in
Fig.~\ref{fig:simulation}. We see that the simulations capture the
line shape in the skyrmion regime quite well and are distinguishable
from the predicted line shapes in the other magnetic phases. Although
the agreement for the single-$q$ helical phase is reasonable, a
discrepancy is evident in the multi-$q$ helical phase, as might be
expected if the trial structure is not the correct one.  (It is possible in this case that an
 alternative trial structure would produce results that match the
 measured spectra more closely.)
We  note that 
our results appear to be consistent with magnetic moments being
localized on the Cu$^{2+}$ 
ions, as was found from x-ray measurements \cite{Langner-2014}
(although it is possible that a delocalized arrangement would match
the data more closely in the multi-$q$ phase).

The simulations may also be used to test whether the muon line-shape
has a sensitivity to the length scale $L_{\mathrm{sk}}$ that
characterizes the SL. Plots of the width $B_{\mathrm{rms}}$ and
skewness $\beta$ against $L_{\mathrm{sk}}$
are shown in Fig.~\ref{fig:simulation}, where we see that there is
little variation with $L_{\mathrm{sk}}$. It would therefore be difficult for this system to
determine the $L_{\mathrm{sk}}$ in the absence of other detailed
information about the phase, such as the magnetic structure, muon
sites and moment sizes. 

\begin{figure}
\begin{center}
\epsfig{file=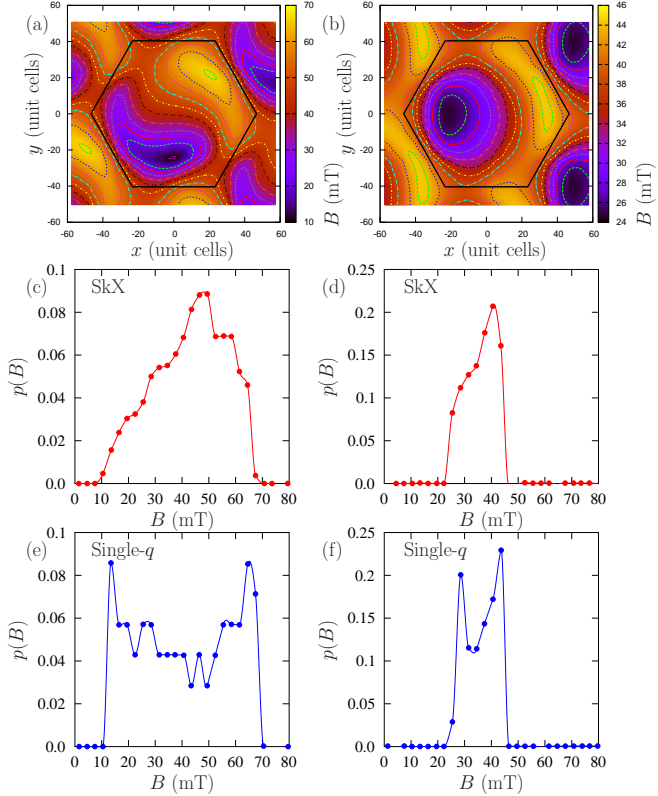,width=\columnwidth}
\caption{The magnitude of the dipole field at the (a) $A$ and (b) $E$ muon
  sites due
  to a skyrmion in an applied field of $B_{\mathrm{a}}=35$~mT. Field distribution for SkX phase at (c) $A$ and (d) $E$
  sites.  Field distribution for the single-$q$ phase at (e) $A$ and
  (f) $E$ sites.  \label{fig:sim2}}
\end{center}
\end{figure}

To gain some insight into how the field distributions shown in
Fig.~\ref{fig:simulation}   arise from the SkX phase spin textures, we
may consider the dipole fields produced by a skyrmion at individual muon
sites. In Cu$_{2}$OSeO$_{3}$ this is complicated by the large number
of inequivalent inequivalent muon sites \cite{Maisuradze-2011}  and the ferrimagnetic
arrangements of Cu$^{2+}$ spins that forms the basis of the skyrmion
spin texture. 
Fig.~\ref{fig:sim2}(a) and (b) show examples of the magnitude of the dipole field found
at two of the five crystallographically distinct muon sites, evaluated across a
single skyrmion. The corresponding field distribution $p(B)$ derived from each is also
shown [Fig.~\ref{fig:sim2}(c) and (d)]. The skyrmion texture is shown to give rise to an asymmetric
field distribution at each site, with spectral weight skewed towards high fields. 
When summed over all muon sites these individual distributions produce 
the total $p(B)$ distribution shown in
Fig.~\ref{fig:simulation}. These distributions may be contrasted with
those
 found in the single-$q$ helical phase  [Fig.~\ref{fig:sim2}(e) and (f)], which are more symmetrical. 
It is difficult from these results to make general predictions regarding
the dipole field distribution expected from other skyrmion materials,
beyond the expectation of asymmetric field distributions at each muon
site. We note that this complexity for the SL spectrum is in marked contrast to the
relative simplicity of the VL spectrum, where characteristic spectral
features map directly onto special positions within the VL.

In conclusion, we have presented results that suggest that TF
$\mu^{+}$SR is sensitive to the
skyrmion lattice via measurements similar to those
carried out on the vortex lattice in type II superconductors. This
would imply that in Cu$_{2}$OSeO$_{3}$ 
the SL is static on a timescale $\tau > 100$~ns. 
We are
also aware of independent work currently being prepared for
publication \cite{liu} that demonstrates that 
longitudinal field $\mu^{+}$SR measurements \cite{hayano} are also sensitive to the
skyrmion phase via changes in the spin dynamics. Taken together, these
results suggest that in favourable cases $\mu^{+}$SR
could be utilized as a method for identifying skyrmionic materials, 
characterizing the behaviour of the skyrmion lattice and deriving
their phase diagrams. 
\acknowledgments

This work was carried out at the STFC ISIS Facility, Rutherford
Appleton Laboratory, UK
and S$\mu$S, Paul Scherrer Institut, Switzerland and we are grateful
for the provision of beam time. 
We thank the EPSRC (UK)
and the John Templeton Foundation for financial support. JCL
acknowledges funding from the Royal Society (UK). 
This work made use of the Hamilton high performance computing (HPC)
cluster based in Durham University and the facilities of the N8 HPC
Centre provided and funded by the N8 consortium and EPSRC (Grant
No.EP/K000225/1). The Centre is co-ordinated by the Universities of
Leeds and Manchester.
We are grateful to  Y.J.\ Uemura and U.K. R\"{o}ssler for useful discussions and Peter Baker (ISIS) and Alex
Amato (S$\mu$S) for experimental assistance. Data presented in this paper will be made available via http://dx.doi.rog/10.15128/kk91fm30h.


\begin{thebibliography}{xx}

\bibitem{chaikin}
P. M. Chaikin and T. C. Lubensky {\it Principles of Condensed Matter
  Physics} (CUP) (1990).

\bibitem{nagaosa}
N. Nagaosa and Y. Tokura, Nature Nano. {\bf 8}, 899 (2013).

\bibitem{skyrm}
 T. H. R. Skyrme, Proc. Roy. Soc. Lond. A {\bf 260}, 127 (1961);
 T. H. R. Skyrme, Nucl. Phys. {\bf 31}, 556 (1962); G. E. Brown and M. Rho (Eds.) {\it The
multifaceted Skyrmion} (World Scientific Singapore) (2010). 

\bibitem{Muhlbauer-2009}   S.   M\"{u}hlbauer,   B.  Binz,   F.  Jonietz,
C. Pfleiderer, A. Rosch, A. Neubauer, R. Georgii, P. B\"{o}ni, Science
{\bf 323}, 915 (2009).

\bibitem{Munzer-2010}   W.    M\"{u}nzer,   A.   Neubauer,    T.   Adams,
S.  M\"{u}hlbauer, C.  Franz,  F. Jonietz,  R.  Georgii, P.  B\"{o}ni,
B. Pedersen, M. Schmidt, A. Rosch and C. Pfleiderer, Phys. Rev. B {\bf
  81}, 041203(R) (2010).

\bibitem{Yu-2010} X. Z. Yu, Y. Onose, N. Kanazawa, J. H. Park, J. H. Han,
Y. Matsui, N. Nagaosa and Y.  Tokura, Nature {\bf 465}, 901 (2010).

\bibitem{Yi-2010a}  X.  Z.   Yu,  N.  Kanazawa,  Y.   Onose,  K.  Kimoto,
W. Z.  Zhang, S. Ishiwata, Y.  Matsui and Y. Tokura,  Nature Mat.
{\bf 10}, 106 (2010).

\bibitem{Seki-2012a} S.  Seki, X.  Z. Yu, S.  Ishiwata and  Y. Tokura,
  Science {\bf 336}, 198 (2012).

\bibitem{Seki-2012b} S.  Seki, J.-H.  Kim, D.  S. Inosov,  R. Georgii,
  B.  Keimer, S.  Ishiwata  and  Y. Tokura,  Phys.  Rev.  B {\bf  85},
  220406(R) (2012).

\bibitem{Seki-2012c} S. Seki,  S. Ishiwata and Y. Tokura,  Phys. Rev. B
  {\bf 86}, 060403 (2012).

\bibitem{Langner-2014}   M.  C.   Langner,  S.   Roy,  S.   K.  Mishra,
  J. C.  T. Lee,  X. W.  Shi, M.  A. Hossain,  Y.-D. Chuang,  S. Seki,
  Y. Tokura, S.  D. Kevan and R. W. Schoenlein,  Phys. Rev. Lett. {\bf 112},
  167202 (2014).

\bibitem{Adams-2012}  T.  Adams,  A.  Chacon,  M.  Wagner,  A.  Bauer,
  G. Brandl,  B. Pedersen, H.  Berger, P. Lemmens, and  C. Pfleiderer,
  Phys. Rev. Lett. {\bf 108}, 237204 (2012).


\bibitem{omrani}
A. A. Omrani, J. S. White, K. Pr\v{s}a, I. \v{Z}ivkovi\'{c}, H. Berger, A. Magrez,
Y.-H. Liu, J. H. Han and H. M. R\o nnow, Phys. Rev. B {\bf 89}, 064406 (2014).

\bibitem{forgan}
M. R. Eskildsen, E. M. Forgan and H. Kawano-Furukawa, 
Rep. Prog. Phys. {\bf 74}, 124504 (2011).


\bibitem{steve}
S. J. Blundell, Contemp. Phys. {\bf 40}, 175 (1999).


\bibitem{sonier}
J. E. Sonier, J. H. Brewer, and R. F. Kiefl, Rev. Mod. Phys. {\bf 72}, 769 (2000).


\bibitem{Bos-2008} J.-W.  G. Bos, C.  V. Colin  and T. T.  M. Palstra,
  Phys. Rev. B {\bf 78}, 094416 (2008).

\bibitem{williams}
D. B. Williams and C.
B. Carter, {\it Transmission Electron
Microscopy} (Springer, New York, 1996), Chap. 28.



\bibitem{WiMDA}
F. L. Pratt, Physica B {\bf 289}, 710 (2000).


\bibitem{Maisuradze-2011}  A.  Maisuradze,  Z. Guguchia,  B.  Graneli,
  H. M.  R{\o}nnow, H.  Berger and  H. Keller, Phys.  Rev B  {\bf 84},
  064433 (2011).


\bibitem{Janson-2014}  O. Janson,  I. Rousochatzakis,  A. A.  Tsirlin,
  M. Belesi, A.  A. Leonov, U.  K.  R\"{o}{\ss}ler, J.  van den Brink,
  H. Rosner, Nature Commun. {\bf 5}, 5376 (2014).




\bibitem{Rossler-2006} U.   K.  R\"{o}{\ss}ler,  A.  N.   Bogdanov and
  C. Pfleiderer, Nature {\bf 442}, 797 (2006).

\bibitem{Petrova-2010} O.  Petrova and O.  Tchernyshyov,  Phys. Rev. B
  {\bf 84}, 214433 (2011).



\bibitem{mochizuki}
M. Mochizuki, X. Z. Yu, S. Seki, N. Kanazawa, W. Koshibae,
J. Zang. M. Mostovoy. Y. Tokura and N. Nagaosa, Nature Materials {\bf
  13}, 241 (2014).

\bibitem{bogdanov}
A. N. Bogdanov, U.K. R\"{o}ssler, C. Pfleiderer, Physica B {\bf 359}
1162, (2005).


\bibitem{liu}
L. Liu,  Paper presented at MuSR2014, Grindelwald, Switzerland
(abstract available from www.psi.ch/musr2014).


\bibitem{hayano}
R. S.  Hayano,
Y. J. Uemura, J. Imazato, N. Nishida, T. Yamazaki and R. Kubo,
 Phys. Rev. B {\bf 20}, 850 (1979).


\end{thebibliography}
\end{document}